\documentclass[useAMS,usenatbib]{mn2e}
\usepackage{amssymb,amsmath}
\usepackage{amsfonts}
\usepackage{epsfig}
\usepackage{graphicx}
\usepackage{ulem}
\input epsf.sty
 \numberwithin{equation}{section}

\newcommand{\bea}{\begin{eqnarray*}}
\newcommand{\eea}{\end{eqnarray*}}
\newcommand{\ba}{\begin{array}}
\newcommand{\ea}{\end{array}}
\newcommand{\p}{\partial}

\newcommand{\ee}{\end{equation}}
\newcommand{\be}{\begin{equation}}

\newcommand{\il}{\int\limits}
\def\d{{\rm d}}
\newcommand{\case}{\textstyle\frac}

\title{Equilibrium models of radially anisotropic spherical stellar systems with
softened central potentials}
\author[E. V. Polyachenko et al.]
       {E.~V.~Polyachenko,$^1$\thanks{E-mail: epolyach@inasan.ru}, V. L. Polyachenko,$^1$
        I. G. Shukhman,$^2$\thanks{E-mail: shukhman@iszf.irk.ru}\\
       $^1$Institute of Astronomy, Russian Academy of Sciences, 48 Pyatnitskya St., Moscow 119017, Russia\\
       $^2$Institute of Solar-Terrestrial Physics, Russian Academy of Sciences,
       Siberian Branch, P.O. Box 291, Irkutsk 664033, Russia}

\date{Accepted \qquad
      Received }

\pagerange{\pageref{firstpage}--\pageref{lastpage}} \pubyear{2011}

\begin{document}
\maketitle

\label{firstpage}

\begin{abstract}
 We study a new class of equilibrium
two-parametric distribution functions of spherical stellar systems with
radially anisotropic velocity distribution of stars. The models are less
singular counterparts of the so called generalized polytropes, widely used
in works on equilibrium and stability of gravitating
systems in the past.  The offered models, unlike the generalized polytropes,
have finite density and potential in the center. The absence of the singularity
is necessary for proper consideration of the radial orbit instability, which is
the most important instability in spherical stellar systems. Comparison of the
main observed parameters (potential, density,
anisotropy) predicted by the present models and other popular equilibrium
models is provided.
\end{abstract}


\section {Introduction}

Equilibrium models of spherical stellar systems are needed for observations
and numerical simulations of open and globular clusters (see, e.g. Kharchenko
et al. 2009, Ernst \& Just 2013). On the other hand, our interest in developing
a new class of radially-anisotropic models is explained by our desire to
perform correct stability analysis of systems with nearly radial orbits. The
gravitational potential and radial force in models of
spherical stellar systems in which all stars travel on purely radial orbits are
singular. This makes it impossible to apply the standard methods of the
linear stability theory, and also cast doubts on some of the works on
Radial Orbits Instability (ROI) (see e.g. Antonov, 1973).

Even a small dispersion in the angular momentum can improve the situation,
however, its presence cannot guarantee the removal of the singularity. An
example is a series of models known as generalized polytropes, which remain
singular despite having some dispersion (see, e.g., Bisnovatyi-Kogan and
Zel'dovich, 1969;  H$\acute{\rm e}$non, 1973). The potential at $r\approx 0$
determines the behavior of the precession rate $\Omega_{\rm pr}$ at small
angular momentum, which plays a significant role in the stability of the system
(Polyachenko et al. 2010). For singular potentials, the precession rate is no
longer proportional to the angular momentum, and very quickly (with infinite
derivative) departs from zero
for angular momentum near $L=0$ (see, e.g., Touma and Tremaine, 1997).
In this case, usual arguments concerning the mechanism of radial orbit instability which,
in particular, involve the linear approximation for the precession rate (see, e.g.,
Palmer 1994) are not useful.

Note that most works that include spectrum determination by matrix methods use
models that cannot be made arbitrarily close to systems with purely radial
orbits. The standard choice is Osipkov-Merritt type DFs (Osipkov, 1979; Merritt
1985).  However, these DFs have restrictions on the largest possible radial
anisotropy.

%%%%%%%%%%%%%%%%%%%%%%%%%

The simplest isotropic self-gravitating polytrope $F(E) \propto (-2E)^q$, where $E =\frac
{1} {2} \, (v_r^2+v _ {\perp} ^2) + \Phi (r) \leq 0$ is the energy  (see,
e.g., Fridman \& Polyachenko 1984) can be used to construct a series of purely radial
models
\begin{align}
F(E) \propto \delta(L^2) (-2E)^q\ ,
\label{eq:pure}
\end{align}
where $\delta(x)$ is the Dirac delta-function, $L=r \, v _ {\perp} $ is the absolute value
of the angular momentum of a star. Generalization of (\ref{eq:pure}) is
possible by replacing the delta-functions on the distribution of the form
$$
\delta(L^2)\ \to\ \frac{H(L^2_T-L^2)}{L_T^2}\ ,
$$
where $H(x)$ is the Heaviside step function. In the limit $L_T\to 0$ the function
$H(L^2_T-L^2)/L_T^2$ becomes the delta-function,
$$\int \frac{\d (L^2)}{L_T^2}\, H(L^2_T-L^2)=1\ ,\ \ \ \lim\limits_{L_T\to
0}\frac{H(L^2_T-L^2)}{L_T^2}=\delta(L^2)\ .
$$

%%%%%%%%%%%%%%%%%%%%%%%%%

The allowed range of parameter $q$ coincides with the range of
the polytropic index in classical polytropic models: $-1\leq q < \case72$ (see,
e.g., Binney \& Tremaine, 2008).
Parameter $L_T$ specifies width of the phase space region over angular momentum
$L$ occupied by the model, $L_T \geq 0$. If $L_T$ is less than some critical
value $ (L_T) _ {\textrm {iso}} (q) $ then radial motions dominate. $ (L_T) _ {\textrm
{iso}} (q) $ has the meaning of the maximum specific angular momentum of the particles
in an isotropic self-gravitating polytrope of index $q$.
For $L_T \geq (L_T) _ {\textrm {iso}} (q) $ models no longer depend on $L_T$ and
become isotropic.

In contrast with the previously used models, the proposed anisotropic
polytropes reach the limit of  purely radial systems for a wide region of
polytropic index $q$.  Besides, relative simplicity of the models allows one to
achieve good accuracy for eigenmodes and stability boundaries, which in turn
can help in understanding the mechanism of ROI.

In Sec. 2 we give general equations and provide profiles of the potential,
density, and anisotropy for the proposed models, and for several other models
commonly used for spherical systems.  Sec. 3 is devoted to the study
properties of the models in the limit $L_T = 0$. Then, in Sec. 4 we explore in more
details several special families of models for which the equilibrium state can
be obtained analytically or stability analysis is particularly simple. Sec. 5
stresses on the orbit's precession behavior of nearly radial orbits, and on
difficulties that arise in systems with purely radial orbits. In Sec. 6 we
summarize the results.

\section {Softened anisotropic polytropes}

In this paper we consider two-parametric series (parameters $q$ and $L_T $)
of models with DF
$$
  F (E, L) = \frac {N} {4\pi^3\,L_T^2} \, H(L^2_T-L^2)\, F_0 (E) \ ,
$$
where $N=N(q,L_T)$ is a constant defined by the normalization condition that the total mass of the
system  $M=1$. For simplicity, we assume that the gravitational constant
and a radius of the spherical system are equal to unity as well: $G=1$, $R=1$. Dependence of the DF
on energy is supposed to be the same as in the classical polytropic models,
\begin {align}
F_0(E)=2\,(1+q)\,(-2 E)^q\ .
  \label {model_E}
\end {align}
The form of (\ref {model_E}) suggests that an additive constant in the
potential $ \Phi_0 (r)$  is chosen in such a way that the potential is equal to
zero on the sphere boundary, $ \Phi_0 (1)=0$. Moreover, the factor $(q+1)$
allows to include the boundary value $q=-1$ in the region of available values,
since  $\lim\limits_{q\to -1^+} F_0(E) = \delta(E)$ (see, e.g., Gelfand and
Shilov, 1964).

Then, it is convenient
to define the relative potential and the relative energy of a star by $ \Psi
(r) =-\Phi_0 (r) \ge 0$, $ {\cal E} =-E \ge 0$, and use the DF in the form:
\begin {align}
 F({\cal E}, L)=\frac{N (q, L_T)(1+q)}{2\pi^3 L_T^2}\,H(L^2_T-L^2)\,(2{\cal E})^q\ .
  \label {model}
\end {align}
 Below, we shall refer to this models as  ``softened'' anisotropic
polytropes or  PPS polytropes.
\medskip

For density distribution one obtains:
\begin{multline}
 \rho (r) = \frac {N} {\pi^2 \, L_T^2} \frac {\Gamma (q+2) \, \Gamma (\frac {3} {2})}
{\Gamma (q +\frac {5} {2})} \,
  (2\Psi) ^ {q+3/2} \\
\times
\left \{\ba {ll} 1\  & \textrm{for } 2\Psi \, r^2 <L_T^2\ , \\
1-\left [1 -{L_T^2} / {(2\Psi r^2)} \right] ^ {q+3/2}& \textrm{for }  2\Psi \,
r^2> L_T^2\ .
 \ea \right.
\label {density}
\end{multline}
Here $ \Gamma(z) $ denotes the Gamma function. With a newly defined function,
\[{\cal F}_n(x) = 1-[1-{\min}(1,x)]^n,\] the expression for density can be
written in the compact form,
\begin{align}
 \rho = A\,(2\Psi)^{q+3/2}\,{\cal F}_{q+\frac{3}{2}}\Bigl(\frac{L_T^2}{2r^2\Psi}\Bigr)\ ,
\end{align}
where
 \[\quad A \equiv \frac {N} {\pi^2 \, L_T^2} \frac {\Gamma (q+2)\,\Gamma
(\frac32)}{\Gamma (q +\frac52)}.
\]

The number of solutions of equation
\begin{align}
 2r^2\Psi(r)  = L_T^2
 \label{eq:r1r2}
\end{align}
depends on the value of $L_T^2$.
Given $L_T$ is less than the
maximum specific angular momentum in the isotropic polytrope,
$$
(L_T) _ {\rm iso} (q) = \max\limits_{r}\,[2\,r^2\Psi(r)]^{1/2} = L_\textrm{circ}(0)\
,
$$
where $L_\textrm{circ}(0)$ is the specific angular momentum of the star with ${\cal E}=0$
in a circular orbit in the polytrope $q$, equation (\ref{eq:r1r2})
has two solutions $0<r_1<r_2<1$.
A condition  $2 r^2 \Psi<L_T^2$  is  satisfied in the regions adjacent to the center, $0
\le r <r_1$,
and to the boundary of the sphere, $r_2 < r \le 1$ (regions I and III
respectively). In the region II ($r_1 <r <r_2$), $2 r^2 \Psi > L_T^2$.
The dependence of $(L_T) _ {\rm iso}
(q) $ is given in Fig. \ref {fig:Liso}{\it a} in Sec. 6.

For $L_T\ll 1$, the solutions $r_1$, $r_2$ tend to 0 and 1, consequently. Since
$\Psi(r_1)  \simeq \Psi(0) + {\cal O}(r_1^2)$, $r_1 \simeq L_T/\sqrt{2\Psi(0)}$.
Similarly, $\Psi(r_2) \simeq -(1-r_2)\Psi'(1) + {\cal O}[(1-r_2)^2]$, so $1-r_2 \simeq
L_T^2/[-2\Psi'(1)] = L_T^2/2$. Here $\Psi'(1)=-GM/R^2=-1$ from $G=R=M=1$.

The Poisson equation
\begin {align}
 \Psi''+\frac{2}{r}\,\Psi'=-4\pi\rho(r)
 \label {pe}
\end {align}
together with boundary conditions:
\begin {align}
\Psi'(0)=0\ , \quad\Psi (1) =0\ , \quad\Psi'(1)=-1
  \label {bc}
\end {align}
determine the potential $\Psi$ and the normalization constant $N = N (q, L_T)$.

Generally, equation (\ref {pe}) with boundary conditions (\ref{bc}) is solved
numerically.  However, several values of $q$ result in analytic solutions. As
is the case for isotropic polytropes, if $L_T>(L_T)_{\rm iso}$, analytic
solutions exist for $q=-\frac{3}{2}, -\frac{1}{2}, \frac{7}{2}$. However, if
$q<-1$, the distribution is unphysical (i.e. unintegrable) and the
$q=-\frac{3}{2}$ case is thus unphysical. The $q=\frac{7}{2}$ case, which
corresponds to the Plummer model, on the other hand results in solutions with
infinite radius, which is also must be rejected due to the boundary condition.
In fact the finite radius condition restricts $q<\frac{7}{2}$. In the  $L_T=0$
limit, analytical solutions are found for $q=-\frac{1}{2}, \frac{1}{2}$, but
the $q=\frac{1}{2}$ case leads to no physically acceptable solutions with
finite radius. Connecting two limits, for arbitrary $L_T$, the equation has
exact analytical solutions if $q=-\frac{1}{2}$, for which the source term of
the equation becomes linear on $\Psi$.  In addition, if $q=1/2$, equation
(\ref{pe}) for region II, where $2\Psi r^2 > L_T^2$,  is analytically solvable.
Hence, in the $q=\frac{1}{2}$ case it is possible to construct approximate
analytic physically acceptable solutions with finite mass and radius, if $L_T >
0$. The solutions for $L_T\ll 1$, which ignore narrow region III (with width
$\propto L_T^2$), are constructed in Appendix.

Fig.\,\ref {fig:psi_rho} illustrates a comparison of the potential and the
density for our model (\ref {model}) at $q =\frac {1} {2} $, $L_T\simeq 0.2$
with the corresponding  profiles obtained for the generalized polytropes
(hereafter GP) (see, e.g., Polyachenko et al., 2011):
\begin {align}
  F_{\rm GP}({\cal E},L)=C(s,q)\,L^{-s}(2\,{\cal E})^q\ ,
  \label{model_op}
\end {align}
and for Osipkov-Merritt  (hereafter OM) models of type
\begin {align}
  F_{\rm OM}({\cal E},L)=A (r_a, p) \, Q^p\ , \quad Q={\cal E}-{\case {1} {2}}\,{L^2}/{r_a^2} \
,
  \label{model_om}
\end {align}
where $A$ is a normalization constant and $r_a $ is the so called anisotropy
radius (Osipkov, 1979; Merritt, 1985).

Parameters of GP and OM models have been
selected in such a way that the global anisotropy (\ref{globalan})
(see below for the definition) in all the models was the same and equal to $ \xi \simeq 0.65 $,
that roughly corresponded to the predominance of the total radial kinetic energy of stars over
the total transversal kinetic energy by factor of $3/2$.
A certain degree of freedom in choosing the parameters $r_a$ and $p$ for the OM DFs
were used to fit the potential and density profiles of the PPS polytrope.

Curves of the potentials for different models almost coincide at $r > r_1 \approx 0.070$.
Difference is noticeable in the central region $r <r_1$, where PPS polytropes become
isotropic.
However, distribution of density is significantly different:
while PPS polytropes and OM models demonstrate similar behavior and  finite density
in the center, the generalized polytropic models show rather strong singularity,
$\rho \sim r^{-s}$.

The \textit{local} anisotropy parameter (see, e.g., Binney \& Tremain, 2008)
\begin {align}
 \beta (r)\equiv 1-{\case {1}{2}}\,{\langle v_\perp^2\rangle}/{\langle v_r^2\rangle} \ ,
  \label {localan}
\end {align}
is an important characteristics of stellar systems. Here $\langle v_{r}^2\rangle$ and $\langle
v_{\perp}^2\rangle$ are dispersions of radial and transversal velocities respectively:
\[
\langle v_r^2 \rangle =\frac{2\pi}{\rho (r)}\int v_r^2 v_\perp \d v_\perp \d
v_r \, F ({\cal E}, L)\]\vspace{-6mm}\[ =\frac{2\pi}{r^3\rho(r)}\int\d{\cal
E}\,\d L^2\,F({\cal E},L)\,(L_{\max}^2-L^2)^{1/2}\ ,\]
\[
\langle v_\perp^2\rangle=\frac{2\pi}{\rho (r)}\int v^3_\perp \d v_\perp \d v_r
\,F({\cal E},L)\]\vspace{-6mm} \[=\frac{2\pi}{r^3\rho (r)}\int\d {\cal E}\,\d
L^2 \, F({\cal E},L)\,L^2\,(L_{\max}^2-L^2)^{-1/2} \ ,
% \label {disp}
\]
where $L_{\max}(r,{\cal E})\equiv \sqrt {2r^2\,[\Psi (r)-{\cal E}\,]
\hspace {-5pt} \phantom {\big |}}$\,.  Either by direct integration
of the DF or using the method of Dejonghe (1986) (see also Dejonghe \&
Merritt, 1992) one obtains for PPS polytropes \[\rho \langle v_r^2
\rangle = \frac{A}{2q+5}\,(2\Psi)^{q+5/2}\,{\cal F}_{q+5/2},\] and
 \begin{align}
\rho \langle v_\perp^2 \rangle = 2 \Psi \rho - (2q+3)\,\rho \langle
v_r^2 \rangle,
 \label {T_T}
 \end{align}
 and so \[\beta = (q+\case52)(1-{\cal
F}_{q+3/2}/{\cal F}_{q+5/2}).\] Profiles of local anisotropy are shown
in Fig.\,\ref {fig:beta}. In the central region I ($r <r_1$) PPS
polytrope is isotropic ($\beta=0$), while beyond this radius (region
II, $r_1 < r$) it quickly becomes radially-anisotropic, $\beta>0$. Note
that in contrast with OM models, anisotropy profiles for PPS polytropes
are non-monotonic. Near the boundary of sphere (region III, $r_2<r<1$)
the velocity distribution again becomes isotropic, and $\beta(r)$
decreases sharply to zero. Experiments with different values of $q$
show that lower $q$ give sharper changes of the anisotropy parameter at
boundaries of regions I-II and II-III, although in general behavior of
$\beta(r)$ changes insignificantly. The anisotropy parameter for
generalized polytropes does not depend on radius, $ \beta =
\frac{1}{2}\,s$.

\begin {figure} % [hbt]
%\onecolumn
\centerline {\hspace {5mm} {\includegraphics [width=80mm] {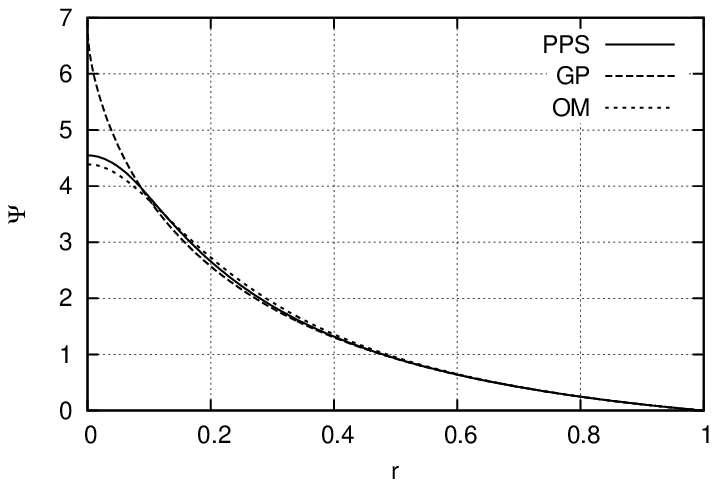}}}
{\vspace{-55mm}a)} \vspace {50mm}

\centerline {\hspace {1mm}{\includegraphics [width=86mm] {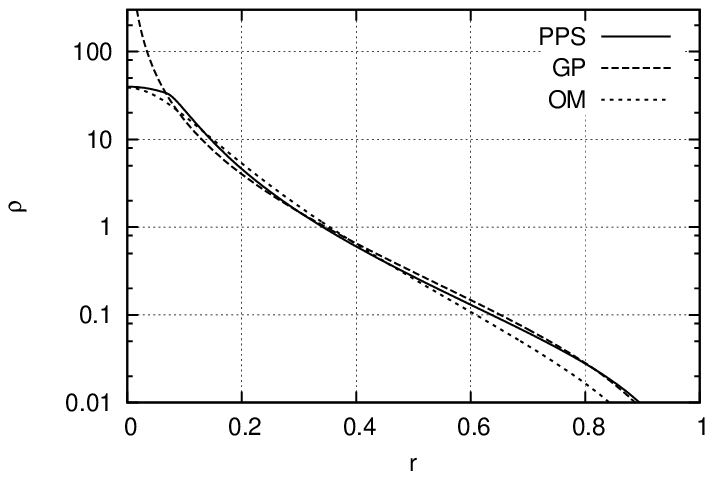}}}
\vspace{-55mm}b) \vspace {50mm}
 \caption {\small (a) Potential $ \Psi (r) $  and (b) density $ \rho (r) $  for PPS
 polytrope, GP for $q=\frac{1}{2}$, and OM model for $p =-1/8$, $r_a=0.12$. Other
parameters of the first two
models ($L_T\simeq 0.2$, $s\simeq 1.3$) were chosen so that global anisotropy
for all models was identical, $ \xi \simeq 0.65$. For PPS polytrope, $r_1 \approx 0.070$,
$r_2 \approx 0.979$.} \label {fig:psi_rho}
 %\twocolumn
\end {figure}

\begin {figure} %[tb]
%\twocolumn
 \centerline {\includegraphics [width=80mm] {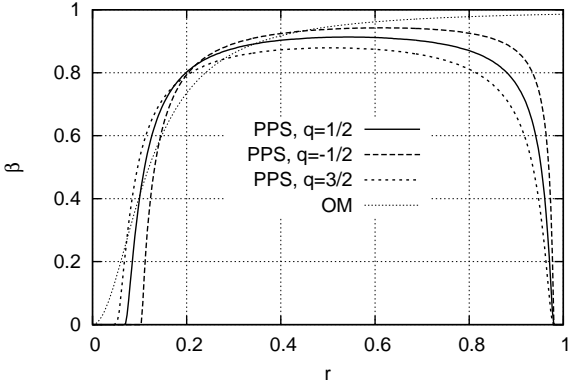}}
\caption {\small  Radial dependence of the local anisotropy $\beta(r)$ for PPS
polytrope, and for OM model,
$\beta(r)=(1+r_a^2/r^2)^{-1} $, for several values of $q$. Parameters of models
are the same as in Fig.\,\ref {fig:psi_rho}. The local anisotropy for classical
GP is constant, $ \beta \simeq 0.65$ (not shown).}
 \label {fig:beta}
\end {figure}

A system as a whole can be characterized by the parameter of {\it global}
anisotropy $\zeta \equiv 2 T_r/T_\perp $ (Fridman \& Polyachenko, 1984), where
$ T_r=4\pi \int_0^1
 \d r\,r^2\rho (r)\cdot{\case{1}{2}}\,{\langle v_r^2\rangle}$ and $ T_{\perp}=4\pi \int_0^1
 \d r\,r^2\rho (r)\cdot{\case{1}{2}}\,{\langle v_{\perp}^2\rangle}$ are total
radial and transversal kinetic energy of all stars in the system.
It is convenient to redefine the anisotropy parameter as follows:
\begin {align}
 \xi (q, L_T) \equiv 1 - {\case {1} {2}} \, {T_\perp} / {T_r} =1-\zeta^{-1}\ .
 \label {globalan}
\end {align}
Then $\xi=0$ corresponds to isotropic systems (in average), while $\xi=1$ implies purely
radial systems. Thus, the definition of $\xi$ is consistent with the definition of
the local parameter $\beta$ and we shall use it henceworth as a global
characteristics for stellar models.

\begin {figure} %[t]
%\onecolumn
 \centerline {\hspace{1mm}{\includegraphics [width=83mm]{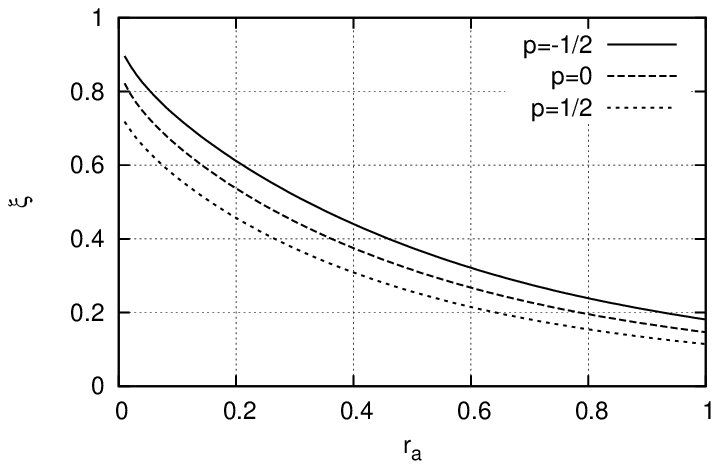}}}
{\vspace{-55mm}a)}\vspace {50mm}

 \centerline{\hspace
{1mm}{\includegraphics [width=83mm] {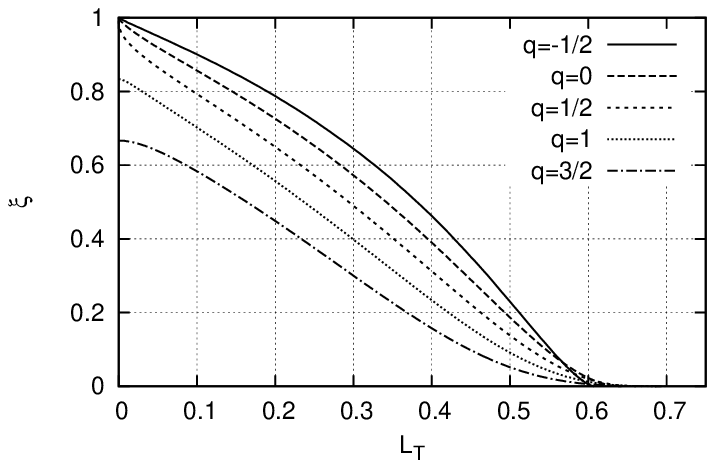}}} \vspace {-55mm}b)
\vspace{50mm}
 \caption {\small Dependence of global anisotropy
(a) for OM model v.s. parameter $r_a$ for $p =-\frac
{1} {2}, \, 0, \, \frac {1} {2} $, and (b) for PPS polytrope v.s. parameter $L_T$
for $q =-\frac{1}{2},\,0,\,\frac{1}{2},\,
1,\,\frac{3}{2}$.}
 \label {fig:xi}
\end {figure}

Comparison of global anisotropy for PPS polytropes and OM model is shown in Fig.\,\ref
{fig:xi}. A characteristic feature of OM model
is that for any parameters $p$ and $r_a$, the value of the
global anisotropy does not reach unity. In contrast, in the PPS polytropes
the limit of purely radial systems exists for a wide range of parameters $q$:
$-1 \leq q <\frac{1}{2}$. This is essential for further study of stability of
systems with nearly radial orbits. Properties of  models near  purely radial
orbits boundary $L_T=0$ are considered in the next section in more details.

For PPS polytropes, from $\langle v_\perp^2 \rangle + (2q + 3) \langle
v_r^2 \rangle = 2\Psi$ (see (\ref{T_T}), one also obtains
\[
%begin{align}
 T_\perp + (2q+3)\, T_r = 4\pi \il_0^R \d r\, r^2 \rho\Psi\ ,
\]
%end{align}
or
\begin{align}
 T_\perp + (2q+3)\,T_r = -\frac{GM^2}R -2W\ ,
\end{align}
where $W$ is the total potential energy of a self-gravitating system
\begin{align}
  W = 4\pi\il_0^R\d r\, r^2\cdot {\case{1}{2}}\,\rho\,\Phi\ ,
  \label{w_xi}
\end{align}
and $\Phi$ is the potential with the zero point given $\Phi(\infty) = 0$; $\Psi
= \Phi(R)-\Phi(r) = -GM/R-\Phi$.  Together with the virial theorem,
$2\,(T_\perp + T_r) + W = 0$ and definition of global anisotropy parameter
(\ref{globalan}), this means that one can express  the total kinetic and
potential energy via $q$ and $\xi$,
\begin{align}
\begin{bmatrix}
 2q+3 &1 &2\\
 2 & 2 & 1\\
 2\,(1-\xi)& -1& 0
\end{bmatrix}
\begin{bmatrix}
 T_r\\
 T_\perp\\
 W
\end{bmatrix} =
\begin{bmatrix}
-GM^2/R\\
 0\\
 0
\end{bmatrix}.
\end{align}
Provided that $\Delta \equiv 7-2q-6\xi \neq 0$,
\[
 T_r = \frac{1}{\Delta}\,\frac{GM^2}{R}\ ,\ T_\perp =
\frac{2\,(1-\xi)}{\Delta}\,\frac{GM^2}{R},\]
 \begin{align}
  W = -\frac{2\,(3-2\xi)}{\Delta}\,\frac{GM^2}{R}\ .
  \label{TTW}
\end{align}
Alternatively, at fixed $q$, the global anisotropy of PPS polytropes is related to the
potential energy:
\begin{align}
 \xi = \frac{(\case72-q)\,w-3}{3w-2}\ ,
  \label{WXI}
\end{align}
where $w \equiv \dfrac{|W|}{GM^2/R} = -\dfrac{W}{GM^2/R}$.

\section {Softened anisotropic polytropes at $L_T \to 0$}

  Specifics of radial and nearly radial systems is a central singularity,
  and therefore they require special consideration.
The GP models (\ref{model_op}) give purely radial orbits at $s=2$. However, not
every $q$ is allowed: as it was noted by H\'{e}non (1973) and Barnes et al.
(1986), no GP exists when $2q+3s\geq 7$. Thus, GPs provide systems consisting
of radial orbits only when $q<1/2$. This also can be seen from our model
equations provided that $L_T=0$. Substituting density
\[
  \rho (r)=\frac{N}{2\pi^2}\,\frac{\Gamma(q+2)\,\Gamma (\frac {1} {2})}
  {\Gamma (q +\frac{3}{2})}\,\frac {(2\Psi)^{q+1/2}}{r^2}
\]
into the Poisson equation and using $x \equiv \ln (1/r) $ as a new independent variable
one obtains
\begin {align}
\frac {d^2\Psi}{dx^2}-\frac{d\Psi}{dx}=-D\,(2\Psi)^{q+1/2}\ ,  \label {pe_x}\\
D\equiv\frac {2N}{\pi}\frac{\Gamma(q+2)\Gamma(\frac{1}{2})}{\Gamma(q+\frac
{3}{2})} \ .
\end {align}
An asymptotic solution for $x\to \infty $ (or for  $r\to 0$) is
\begin {align}
  \Psi(x) \propto x^{m}\ , \quad m=({\case{1}{2}}-q)^{-1}>0\ ,
  \label {rad_psi}
\end {align}
from where we infer that such solutions are
possible for $q <\frac {1}{2}$ only.

For $ q = \frac{1}{2} $, equation (\ref{pe_x}) becomes linear and has exact analytical
solutions. Unfortunately, from it's two linearly independent solutions it is impossible to
construct a solution which would have a finite mass and finite potential energy. However,
if we admit arbitrarily small smearing, $ L_T \ne 0 $, a solution with a finite radius is
possible (see Appendix for details).

The models with purely radial orbits are always singular,
and the singularity is not weaker than $ \rho \propto r^{-2} $. This was first pointed out
by Bouvier \& Janin (1968) (see also Richstone \& Tremaine, 1984). However, it is
more accurate to say that
the singularity may be slightly stronger or slightly weaker than $ r^{-2}$: $ \rho(r)
\propto
r^{-2} [\Psi(r)]^{1/2 + q} $. Since $\Psi(r)\propto[\,\ln(1/r)]^{m} $
with positive $m$  [see (\ref{rad_psi})] one obtains that,
for $q <-\frac{1}{2}$ the singularity is slightly weaker than $r^{-2}$.

%%%%%%%%%%%%%%%%%%%%%%%%%%%%%%%%%%%%%

In the limit of $L_T\to 0^+$, asymptotic
solution for $ q> \frac12 $ takes the form
$\Psi(x) \propto \exp\,(x)=1/r-1$, i.e. models degenerate into a point
(considering the adopted length unit). The normalization constant in this case
tends to zero: $N (q, L_T) \propto L_T^{2q-1} $ at $L_T \to 0$.
Global anisotropy $\xi $ for these models is less than one, which is evident, e.g., from
Fig.\,\ref{fig:xi}\,b. It may seem that there is a contradiction: on one hand
the parameter
$L_T$ tends to zero, and on the other hand the parameter $\xi$, which
characterizes the anisotropy
of the system as a whole, tends to a finite limit less than one. In reality, of course, there is no
contradiction.
With an increase of polytropic index $ q $ the number of particles with energy $ E \sim $ 0
decreases, and the particles with energies close to the minimum potential energy begin to dominate.
For small $ L_T $, the potential well near the center is very deep, so the mass is concentrated
near the center in a very small region of $ r \lesssim {\cal O} (L_T^2) $. Outside this region,
the potential is actually Keplerian, $ \Psi (r) = 1/r-1 $. In fact, radius $ r = 1 $ is
infinitely remote from the region of localization of the mass.

To determine the shape of orbits trapped in this region, one should not rely only
on the smallness of the angular momentum in units $(GMR)^{1/2}$. For highly
elongated orbits,
the angular momentum $L$ should be small compared to an
angular momentum of a circular orbit of the same energy $L_{\textrm
{circ}}(E)$, i.e. $L/L_{\textrm {circ}} (E) \ll 1$.
In other words, when $L_T$ is small compared to one, orbits must not
be nearly radial, and anisotropy parameter $\xi $ is not required to be close to unity.

\begin {figure}%[t]
\centerline {\hspace {0mm} {\includegraphics [width=88mm] {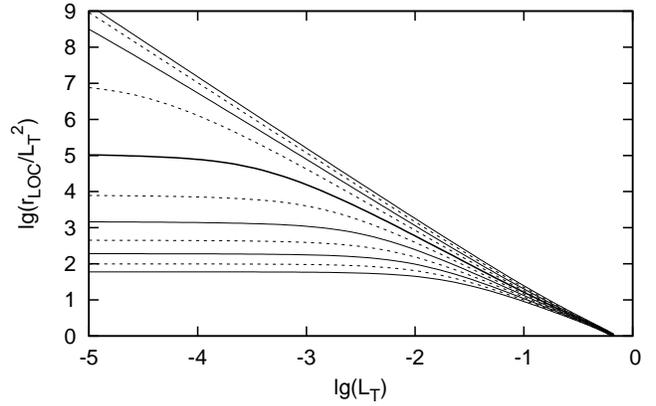}}} \vspace
{0mm}
 \caption {\small  Dependence of localization radius $r_{\rm LOC}$ on $L_T$ at $q>\frac{1}{2}$.
 Eleven curves are shown, starting from $q=0.5$ with step 0.05 (from top to bottom). It is seen,
that beginning from the fifth curve ($q=0.7$, heavy solid line) the curves
 $r_{\rm LOC}(q,L_T)/L_T^2$ tend to constant values for $L_T\lesssim 10^{-5}$.}
 \label {fig:r_loc}
\end {figure}

To illustrate this we define a localization radius $r_{\rm LOC}$ by
the equation
 \begin {align}
  \left[\frac{d \ln\rho(r)}{d\ln (1/r)}\right]_{r=r_{\rm LOC}}=3\ ,
 \label {r_loc0}
\end {align}
which is the radius where the density begins to decrease more rapidly than
$r^{-3}$. The reason is that beyond this radius the gravitational force is
determined primarily by the mass confined withing $r_{\rm LOC}$. From
Fig.\,\ref{fig:r_loc} it is seen that models with $q=0.7$ (fifth curve from
above given by heavy solid line) tend to it's asymptotics $r_{\rm LOC}\propto
L_T^2$ already for $L_T \sim 10^{-5}$. In fact, this behavior occurs for all
values of $q>0.5$, but in order to demonstrate this, we must consider $ L_T$
orders of magnitude less than $L_T\sim 10^{-5}$, which is difficult to
implement numerically.

 For $L_T \to 0^+$, the global anisotropy $ \xi $ as a function of $q$
can be obtained analytically, if $q>\frac {1}{2}$. The Keplerian
potential of the system corresponds to a point mass, if the outer
boundary $R$ is finite. On the other hand, if the system is scaled so
that the potential in the center is finite, then $R\to\infty$, and
`surface term' $GM^2/R$ in (\ref{TTW}) and (\ref{WXI}) becomes zero.
Since all energies cannot all together vanish, it requires the
determinant $\Delta=0$, i.e.
\begin {align}
\xi = {\case{1}{3}}\,({\case {7} {2}}-q)\ .
\label {lin_xi}
\end {align}
We see that when parameter $q$ varies from $\frac{1}{2} $ to $\frac{7}{2}$ the models are
transformed from a model with purely radial orbits to an isotropic one with $ \xi = 0$.

Note that for $ q \to \frac {7}{2} $, the equation (\ref{pe}) reduces to the
Lane-Emden equation for any finite value $ L_T\gg \delta^{1/2} $, where
$\delta\equiv\frac{7}{2}-q\ll 1$. Indeed, introducing variables $\Psi =
\psi/\delta $, $r = z\,\delta$, $N=n_0\,\delta^2\,L_T^2$, we can express
(\ref{pe}) in the form
\[
 \frac{\d^2\psi}{\d z^2}+\frac{2}{z}\,\frac{\d\psi}{\d z}=-{\case{63}{4}}\,n_0\,
\psi^{5} {\cal F}_5\Bigl(\frac{L_T^2}{2\delta\psi z^2}\Bigr)
\]
with boundary conditions:
\[
\psi (1/\delta) =0\ , \  \psi ' (1/\delta)
=-\delta^2\ , \ \psi'(0)=0
\]
which can be replaced by homogeneous boundary conditions
at the origin and at infinity. Point $z_1$ at which 2$ \psi (z_1) \, z_1^2 \approx
2 z_1=L_T^2/\delta $ also goes to infinity, provided that $L_T^2/\delta\gg
1$. The result is the Lane-Emden equation
\[
\psi''+\frac{2}{z}\,\psi' =-{\case{63}{4}} \,n_0\,\psi^5
\]
the solution of which gives the well-known Plummer potential
$\psi=(a^2+z^2)^{-1/2}$ with $a =\sqrt{\frac{21}{4}\,n_0}$, which corresponds
to the isotropic polytropic model with $q=\frac{7}{2}$. Our calculations give
 $n_0\approx 0.00183$, i.e, $a \approx 0.098$.

\section {Special families}

Here we consider several special families of PPS polytropes for which the equilibrium
state can be obtained analytically or stability analysis is particularly
simple: $q=\frac{1}{2}$, $q=0$, $q =-\frac{1}{2}$, $q=-1$.

\subsection {Models with $q=\frac{1}{2}$}

The model with a DF
\[
 F (E,L)=\dfrac {3N}{4\pi^3}\,\dfrac{H(L^2_T-L^2)}{L_T^2}\,\sqrt{-2E}
\]
is a boundary model, which in the limit $L_T\to 0^+$  is turned into purely
radial one, i.e. $\xi(\frac{1}{2},0^+) = 1$, see Fig.\,\ref {fig:xi}b.

\begin {figure} %[tb]
\centerline {\hspace {2mm} {\includegraphics [width=85mm] {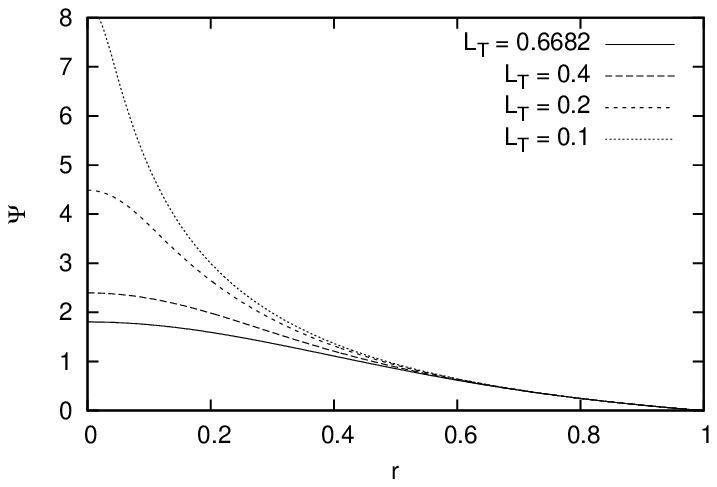}}}
{\vspace{-55mm} a)}\vspace{50mm}

\centerline{\hspace{-2mm}{\includegraphics [width=91mm] {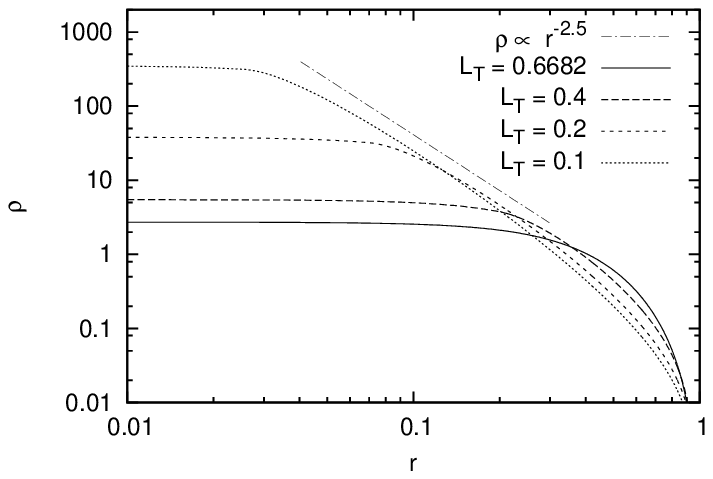}}}
 {\vspace {-55mm} b)} \vspace {50mm}
  \caption {\small (a) The potential profiles and (b) the density profiles for models with
$q=\frac{1}{2}$ and several values of $L_T$. The maximum value of $L_T$ plotted
corresponds to
$(L_T)_{\textrm{iso}}(q)$ and so the corresponding model is identical to the isotropic
polytrope of index $q$.  The dash-dotted line shows the density slope
$\rho \propto r^{-2.5}$.}
 \label {fig:gen_qp05}
\end {figure}

Designation ``$0^+$'' emphasizes the already mentioned fact that for
$q=\frac{1}{2}$ there is no physically acceptable model with a purely radial
orbits, although models with arbitrarily small but finite angular momentum
dispersion are possible. Solving the Poisson equation (\ref {pe}) with density
given by (\ref {density}), it is possible to obtain potential and density
profiles for different $L_T$ in the range $0 < L_T < 0.6682$ (see Fig.\,\ref
{fig:gen_qp05}). It turns out that for small values $L_T$ it is possible even
to obtain analytical expressions for the potential, density and the
normalization constant $N$. The details of this solution are described in
Appendix.

\subsection { ``Step'' models, $q=0$}

 The simplest anisotropic model allowing  both energy and angular momentum
to vary in finite intervals corresponds to parameter $q = 0$:
\[
 F (E, L)=\dfrac{N}{2\pi^3}\,\dfrac{H (L^2_T-L^2)}{L_T^2}\,H(-2E) \ .
\]
Study of the stability of such a DF is the simplest, and at the same time, the
model is quite realistic.

\begin {figure}% [tb]
\centerline {\hspace {4mm}{\includegraphics [width=85mm] {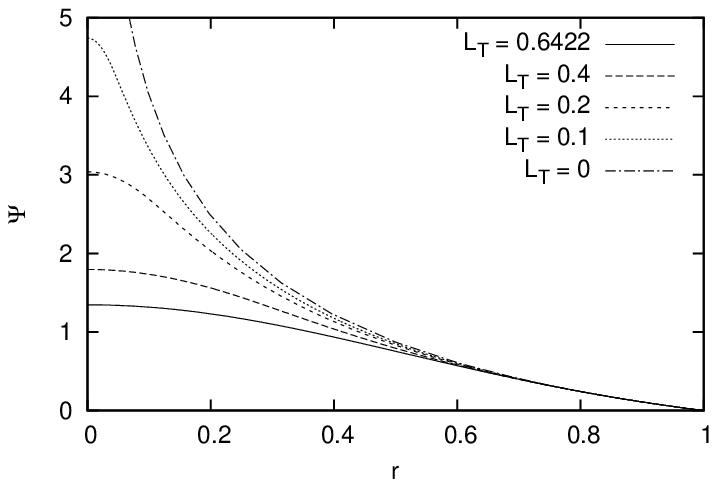}}}
{\vspace {-55mm} a)} \vspace{50mm}

\centerline{\hspace{0mm}{\includegraphics [width=90mm] {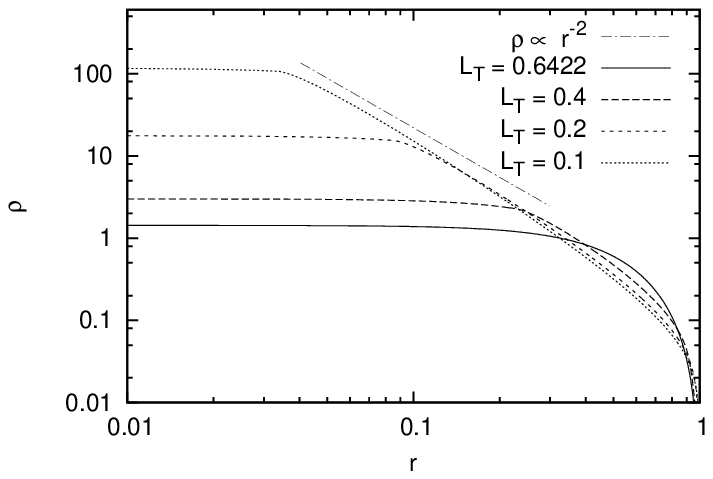}}}
{\vspace {-55mm} b)} \vspace {50mm}
 \caption {\small Same as in Fig.\,\ref{fig:gen_qp05} for $q=0$. The dash-dotted line
shows the density slope $\rho \propto r^{-2}$.}
 \label {fig:gen_q0}
\end {figure}

Solving the Poisson equation (\ref{pe}) it is possible to obtain profiles of the potential and
density for different values of $L_T$ in the range $0 <L_T <0.6422$ (see the
Fig.\,\ref{fig:gen_q0}).

Fig.\,\ref {fig:gen_q0}b demonstrates transformation of density profiles with decreasing $L_T$.
The model with purely radial orbits has a cuspy profile $ \rho \sim \ln (1/r)/r^2$.
Density profiles of nearly radial models differ from the cuspy profile only in a small region near
the center $r <r_1 \sim L_T $.

Dependence of global anisotropy $\xi(L_T) $ for this model is presented in
Fig.\,\ref {fig:xi}. It is seen, that the limit $L_T \to 0$ exists and the global
anisotropy tends to one.

\subsection {Models with $q =-\frac{1}{2}$}

This anisotropic model is of interest since it allows the exact analytical
solution of the Poisson equation. For $q =-\frac{1}{2}$, the expression for
density can be simplified
\[
 \rho (r) = \frac {N} {4\pi \, L_T^2} \,
\left \{\ba {ll} 2\Psi&\textrm{for }  2\Psi \, r^2 <L_T^2\ , \\
& \\
L_T^2/r^2 &\textrm{for } 2\Psi\,r^2> L_T^2\ .
 \ea \right.
\]
As it was discussed above, in general, there are three regions separated
by radii $r_1$ and $r_2$ ($r_1 <r_2$).
Taking into account boundary conditions (\ref{bc}), the
potential can be written in the form
\[
 \Psi (r) = \left \{\ba {ll}
\Psi_I (r) \equiv A \, \dfrac {\sin kr} {r}\ , & r <r_1\ , \\
\Psi _ {II} (r) \equiv -N \, \ln r+C_1 +\dfrac {C_2} {r}\ , & r_1 <r <r_2\ , \\
\Psi _ {III} (r) \equiv \dfrac {\sin\,[k\,(1-r)]} {kr}\ ,&r_2 <r <1\ ,
 \ea \right.
\]
where $k^2 = 2N/L_T^2$. To find six unknowns $C_1$, $C_2$, $A$, $N$, $r_1$
and $r_2$ there is a set of 6 algebraic equations: 4 conditions of continuity
of the potential and it's first derivative at points $r_1$ and $r_2$, and 2
conditions (\ref{eq:r1r2}) for determining the positions of $r_1$ and $r_2$.

\begin {figure} %[t!]
\centerline {\hspace {0mm}{\includegraphics [width=85mm] {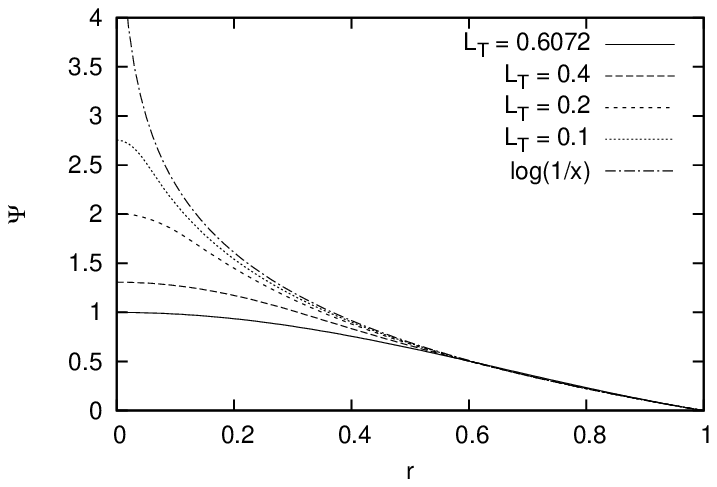}}}
{\vspace {-55mm} a)} \vspace{50mm}

\centerline{\hspace{0mm}{\includegraphics [width=89mm]{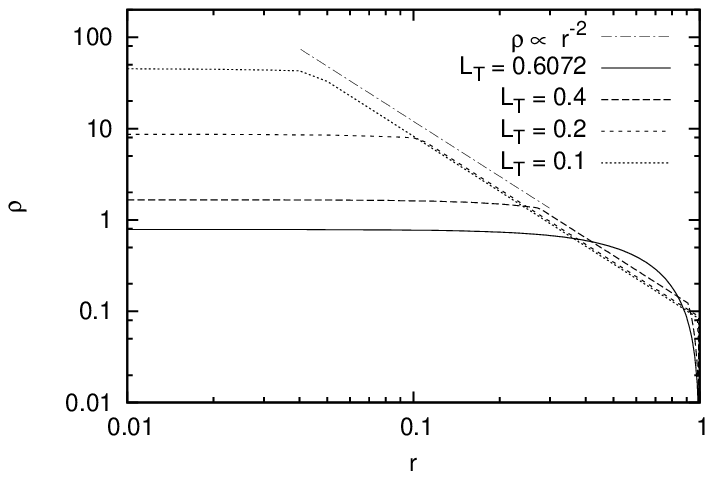}}}
\vspace {-55mm} b) \vspace {50mm}
 \caption {\small Same as in Fig.\,\ref{fig:gen_q0} for  $q=-\frac{1}{2}$.}
 \label {fig:gen_q05}
\end {figure}

Table\,\ref{tab1} gives the solutions of the model parameters for several
values of $L_T$. Corresponding potential and density profiles are shown the in
Fig.\,\ref {fig:gen_q05}. Note that in the purely radial model the density does
not vanish on the boundary $r=1$. In this case $\rho = 1/(4\pi r^2)$, $N=1$,
i.e., radial  dependence of density is the same as that of the isothermal
polytropic model.

\begin {table} %[tb]
\setlength {\tabcolsep} {6pt}
\centering \footnotesize
\begin{tabular}{|c|c|c|c|c|c|c|}
  \hline
$L_T$ & $N$ & $r_1$ & $r_2$ & $A$ & $C_1$ & $C_2$ \\
\hline
  0.6072 & 1.8194 & 0.6458 & 0.6458 & 0.3183 & 0.5817 & -0.6039 \\
  \hline

0.4000 & 1.2784 & 0.2747 & 0.9102 & 0.3270 & 0.2218 &-0.2235 \\
\hline
0.2000 & 1.0854 & 0.1051 & 0.9795 & 0.2723 & 0.0743 &-0.0744 \\
\hline
0.1000 & 1.0325 & 0.0441 & 0.9950 & 0.1917 & 0.0299 &-0.0299 \\
\hline
 \end {tabular}
\caption {Solutions for the parameters of the potentials for $q=-\frac{1}{2}$
and several values of $L_T$.} \label {tab1}
\end {table}

\subsection {Models with $q =-1$}

In the limit $q\to-1$ the PPS polytropes turn into mono-energetic models
\begin {align}
 F (E, L_T) &= \frac {N} {4\pi^3 L_T^2} \, H (L^2_T-L^2) \, \delta (E)\ , \label
{medf}
\end {align}
with density
\begin {align}
 \rho (r)=\frac {N}{\pi^2\,L_T^2}\,(2\Psi)^{1/2}\,{\cal F}_{\frac{1}{2}}\Bigl(\frac{L_T^2}{2\,r^2\Psi}\Bigr).
\label {medens}
\end {align}

\begin {figure}% [tb]
\centerline{\hspace{0mm}{\includegraphics [width=87mm]{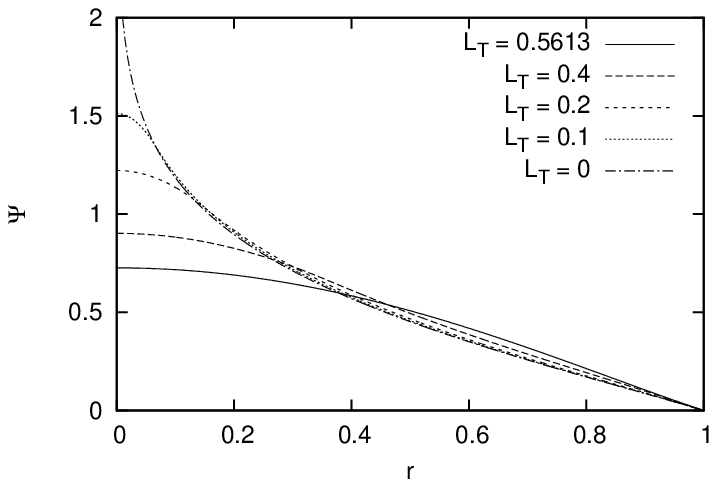}}}
{\vspace{-55mm}a)} \vspace{50mm}

\centerline{\hspace{0mm}{\includegraphics [width=87mm] {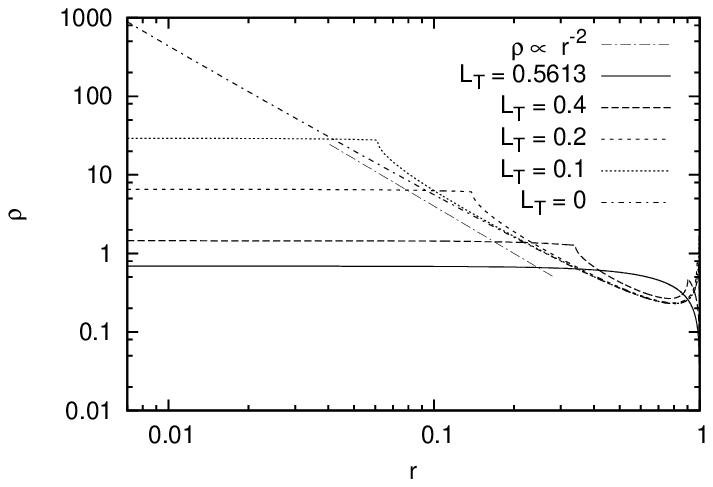}}}
{\vspace{-55mm} b)} \vspace{50mm}
 \caption {\small Same as in Fig.\,\ref{fig:gen_q0} for  $q =-1$.}
 \label {fig:gen_q1}
\end {figure}

Fig.\,\ref{fig:gen_q1} shows the profiles of the potential and density for $q =-1$ and
several values of $L_T$. The potential profiles are monotonic for
all values of parameter $L_T$. At $L_T =0$ the potential has
a central singularity $\Psi \propto [\, \ln (1/r)] ^ {2/3} $, in agreement with
the earlier obtained expression (\ref{rad_psi}) (see also Agekyan, 1962). On the
contrary, density profiles
appear to be non-monotonic, except for the case of isotropic model $L_T = L_\textrm
{iso}$.

\section {The precession of orbits}

In this section we discuss precession of orbits and emphasize related problems
that arise in systems with purely radial orbits with an example of models with
$q=-\frac{1}{2}$. Such a choice is determined by the availability of an
analytical
expression for the potential for the purely radial system in this family, which
is $ \Psi =-\ln r $.

A star azimuth gains a rotation angle $ \Delta\varphi $ during one radial period:
\[
\Delta\varphi=2L\,\int\limits_{r_{\rm min}}^{r_{\rm max}}\frac{\d r}{r^2 \,
\sqrt{2\,(E-\ln r)-{L^2}/{r^2}\!\!\phantom{\big |}}} \ .
\]
Let $ \alpha \equiv L/L_{\rm circ}(E)$ be a ratio of the angular momentum $L$ to the angular
momentum of a star on the circular orbit with the same energy $E$,
$L_{\rm circ}(E)=\exp\,(E-\frac{1}{2})$. Changing the integration variable from
$r $ to $x\equiv r\,\exp\,(-E)$, we obtain (see also Touma and Tremain, 1997):
\begin {align}
g(\alpha) \equiv \Delta\varphi =\frac {2\alpha}{\sqrt {{\rm e}}}\int\frac {\d x} {x
\,\sqrt{-2x^2\ln x-{\alpha^2}/{\rm e}\!\!\phantom{\big |}}} \ , \label {eq:g1}
\end {align}
where ${\rm e}=\exp(1)$.  Note that in variables $ (E, \alpha) $ the
rotation angle is independent of energy. This is the case in all scale-free
potentials such as $\Phi = K\, r^n$, or $\Phi=K\,\ln r $. The integration in
(\ref {eq:g1}) is over all $x$ for which the radicand is positive.  An explicit
expression for function $ g(\alpha) $ and its asymptotic expansion for nearly
radial orbits can be obtained (using the Mellin transform).
After some manipulations, one finally arrives at:
\[
  g (\alpha) = \pi+\frac {1}{\sqrt {\pi}} \,{\rm p.v.}\! \int\limits_0 ^ {\infty}
 \!\!\alpha^t (2{\rm e})^{-{t}/{2}}\sin ({\case {1}{2}}
  \, \pi t)\,t^{{t}/{2}-1}\,\Gamma\Bigl(\case{1}{2}\,(1-t)\Bigr)\,\d t,
\]
where `p.v.' stands for the principal value. Its asymptotic expansion at small
$ \alpha $ is:
\[
  g (\alpha) = \pi +\case {1} {2} \, \pi\mu \, \bigl (1 +\case {1} {2} \, \mu \,
  \ln 2\mu\bigr) + {\cal O} \left (\mu^3\ln^2\mu\right)\ ,
\]
where
\[
  \mu =\frac{1}{\ln\left(1/\alpha\right)}\ .
\]
The precession rate $\Omega_{\textrm {pr}}$ is expressed through $g (\alpha)$ using the relation:
\[
  \Omega_{\textrm {pr}}=\Omega_2-{\case{1}{2}}\,{\Omega_1}=\frac {\Omega_1}
  {2\pi}\, \bigl [g (\alpha)-\-\pi\bigr]\ ,
\]
where $ \Omega_{1,2}(E,L)$ are radial and azimuthal frequencies
\[
  \frac{1}{\Omega_1}=\frac{1}{\pi}\int\limits_{r_{\rm min}}^{r_{\rm max}}\frac{\d r}{\sqrt{2E+2\Psi
  (r)-L^2/r^2\!\!\phantom{\big|}}}\ ,
\]\vspace{-3mm}
\[
  \frac {\Omega_2} {\Omega_1} = \frac {L} {\pi} \int\limits_{r_{\rm min}}^{r_{\rm max}}
  \frac{\d r}{r^2\,\sqrt{2E+2\Psi(r)-L^2/r^2\!\!\phantom{\big|}}}\ ,
\]

For nearly radial orbits $\alpha\ll 1$, we have $\mu\ll 1$ and
$\Omega_1(E,L)\approx\Omega_1(E,0)=\sqrt{{2}/{\pi}}\,\exp(-E)$, so that the
precession rate is
\[
  \Omega_{\textrm {pr}}\approx \frac{1}{\sqrt{8\pi}}\,\mu\,\bigl (1
  +\case{1}{2}\,\mu\,\ln 2\mu\bigr)\,e^{-E}\ .
\]

\begin {figure}% [bt]
\centerline {\hspace {2mm}{\includegraphics [width=87mm] {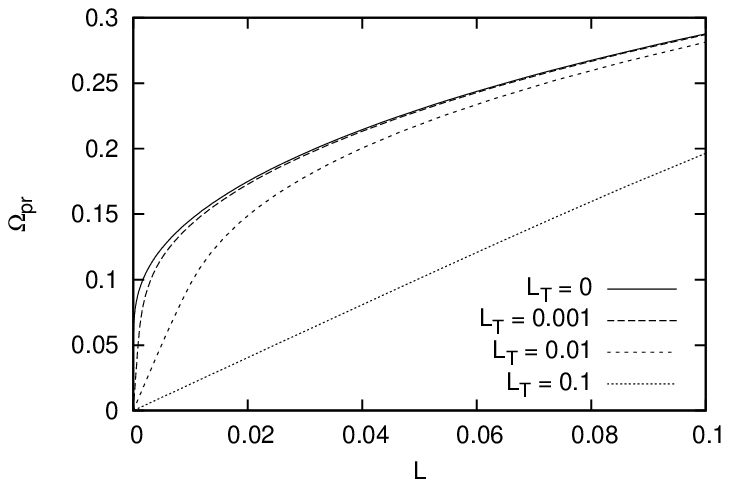}}}
{\vspace{-55mm} a)} \vspace{50mm}

\centerline{\hspace{2mm}{\includegraphics [width=87mm] {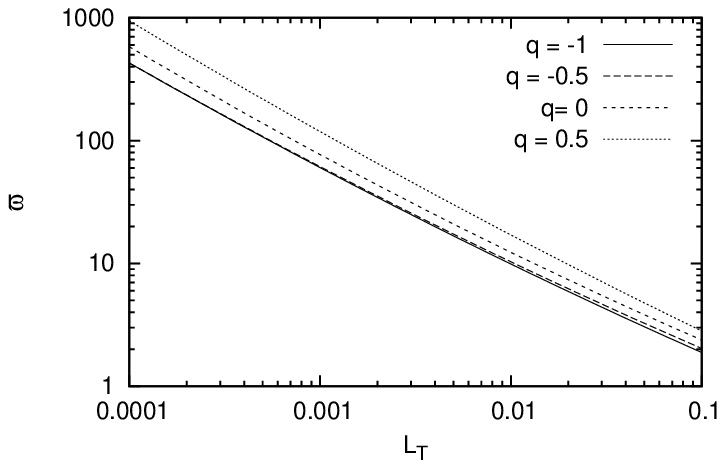}}}
 \vspace{-55mm} b) \vspace {50mm}
  \caption {\small (a) Dependence of the precession rate $ \Omega_\textrm {pr} (E=0,
  L) $ for $q =-\frac{1}{2}$ and several values $L_T$ for PPS polytropes. (b)
Profiles of the precession rate slopes $\varpi(E=0)$ v.s. parameter $L_T$ for
several values of
  $q$.}
  \label{fig:ompr}
\end {figure}

The profiles $\Omega_{\textrm {pr}}(L)$ for several nearly radial systems are
shown in Fig.\,\ref{fig:ompr}a. It is seen that  precession rates depart
quickly from zero at $L=0$, and the slope is steeper for models with lower
$L_T$. Thus, the derivative $ \varpi (E) \equiv \bigl[\p\Omega_{\rm pr}/\p
L\bigr]_{L=0}$ tends to infinity as $L_T \to 0$.

This anomaly is quite typical for highly anisotropic models (including ones
composed of purely radial orbits) in the class of  GP, $F\propto (-2E)^q
L^{-s}$. Since all of these models have gravitational force $\Psi' \propto
r^{1-s}$ near the center (H\'{e}non, 1973), it is singular for highly
anysotropic DFs with $s>1$. In Fig.\,\ref{fig:ompr}b the profiles $\varpi(E)$
v.s. parameter $L_T$ for different polytropic indices $q$ are shown.

\section {Conclusion}

\begin {figure}% [t]
\centerline {\hspace {0mm} {\includegraphics [width=80mm] {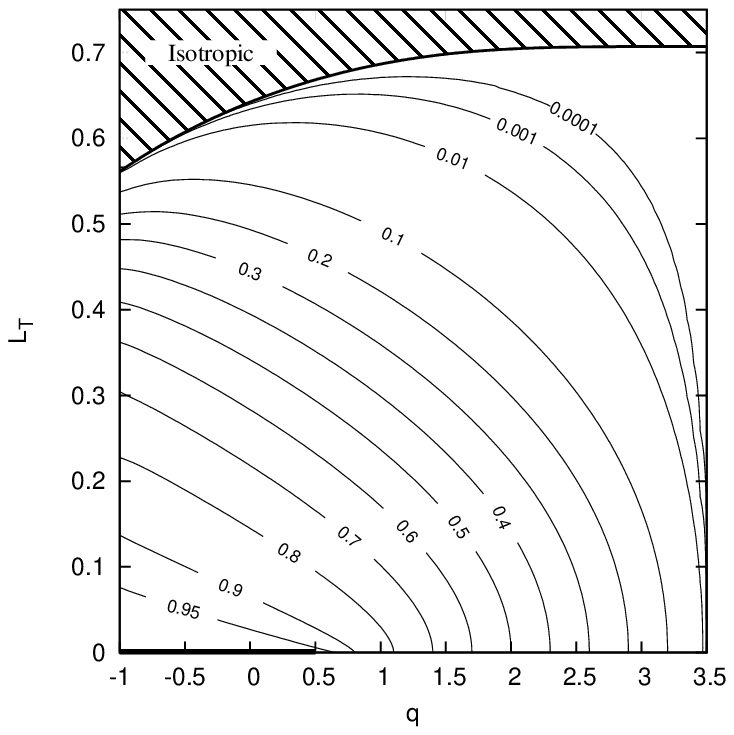}}}
\vspace{-78mm}a)\vspace{74mm}

\centerline{\hspace{2mm}{\includegraphics [width=80mm] {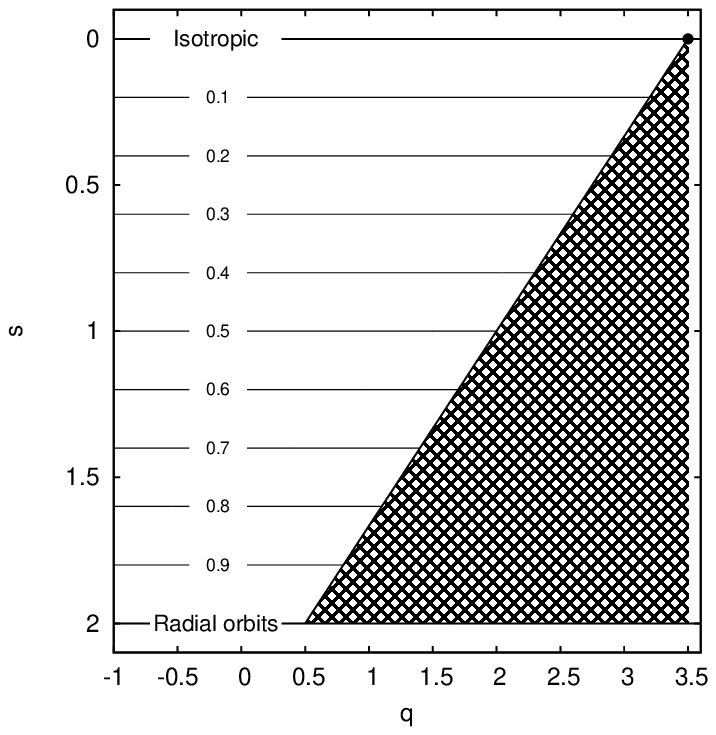}}}
 \vspace{-78mm} b) \vspace {74mm}
 \caption {(a)
Dependence of critical angular momentum $(L_T)_{\rm iso}(q)$ (heavy line) and
isolines of the global anisotropy $\xi$ in the domain ($q, L_T$) of PPS
polytropes. A part of the $x$-axis marked by a heavy line shows the models with
purely radial orbits. (b) Domain of the parameters and isolines of anisotropy
parameter $\xi$ for GP.}
 \label {fig:Liso}
\end {figure}

In this paper we proposed and studied two-parameter models of anisotropic
spherical stellar
systems. Dependence of DFs $F(E,L)$ on the energy $E$ is adopted from the polytropic and generalized
polytropic models. The dependence on the angular momentum is chosen in the form of the Heaviside
function $H(L^2_T-L^2)$, that allows only stars with the angular momenta $L <L_T $. For a
given value of the polytropic index $q$, there is some critical value $L_\textrm{iso}$ of $L_T$,
above which the DFs are ergodic, and the systems are isotropic (see
Fig.\,\ref{fig:Liso}). The curve
$L_T = L_\textrm{iso}(q)$ determines the upper boundary for the model parameters in ($q,L_T$)-plane.

The left and right boundaries of the permissible parameters coincide with the boundaries of the
polytropic models. The left boundary is $q=-1$, where all stars have the same zero energy. The
right boundary is a straight line $q=\frac72$, where the models degenerate
into the
Plummer model and become isotropic for all values ​​of $L_T$. There is no homogeneous model
(one with the density independent of radius), because the corresponding value $q=-\frac32$
is outside the permissible interval.

A natural lower boundary for possible model parameters is the horizontal axis $ L_T = 0 $. However,
not all of the models with $L_T=0$ are purely radial systems. Recall that purely radial models
are models for which the global anisotropy parameter $ \xi=1 $ (see (\ref {globalan})).
Fig.\,\ref {fig:Liso}a shows isolines  $\xi(q, L_T)={\rm const}$ in the model's domain. The
isotropic models correspond to $\xi(q,L_T)=0$.

The most important feature of the proposed models is the existence of a wide
region for parameter $q$, $ -1\leq q <\frac12 $, for which the limit  $L_T=0$
means the purely radial systems. This will enable us to use them in consistent
analytic and numerical study of ROI, which cannot be performed correctly using
systems with purely radial orbits only.

Outside this range, $q>\frac12$, $L_T\to 0^+$ the potential degenerates into
the Keplerian one, $ \Psi (r) = 1/r-1$, and models turn into points. We show
that the global anisotropy $\xi (q,0^+) $ varies linearly with polytropic index
$q$ (see (\ref{lin_xi})) from 1 to 0, which corresponds to transformation of
models from purely radial to the isotropic ones.

Comparison of the parameter domains of PPS polytropes and GP is
possible, recalling the relation $\xi=s/2$ for GP.  In $(q,s)$-plane
the boundary is a trapezoid with a vertical straight line $q =-1$, two
horizontal straight lines $s=0$ and $s=2$ and a sloping side $2q+3s=7$
(H\'{e}non 1973, Barnes, etc. 1986) or $2q+6\xi=7$ (see Fig.\,\ref
{fig:Liso}b). The straight line $s=2$ corresponds to part of the
boundary $q <\frac12$, $L_T=0$, and the sloping side corresponds to
another part of $x$-axis: $q> \frac12$, $L_T\to 0^+ $. The right
boundary of our model $q =\frac{7}{2}$ corresponds to single point
$\bigl(q =\frac{7}{2},\ s=0\bigr)$ in the domain for the generalized
polytropes (the Plummer model). Thus, the domain boundary for PPS
polytropes coincide with the domain boundary of GP. This is not
surprising, since if $q>\frac{1}{2},\ L_T=0^+$ the mass of the system
is localized near the center. In fact, it means that sphere radius $R$
tends to infinity. But just the same $R \to \infty$  occurs when
reaching the boundary $2q=3s=7$ in GP (see H\'{e}non 1973).

For a fixed $L_T $, central density concentration grows with
increasing of the polytropic index $q$. For $q <-\frac{1}{2}$, the anisotropic
models have intervals of growing density at the periphery of spheres. The
Agekyan's model (1962) which is a particular case of our series at $q =-1$,
$L_T=0$, also has this feature.

For the model $q=-\frac12$, we consider the precession rates at low angular
momenta for nearly radial and purely radial orbits. Features of its behavior
play a significant role for stability, first of all in the emergence of ROI
(Polyachenko, etc. 2011). We have shown that in the limit of the purely radial
systems, the derivative of the precession rate over $L$ at $L=0$ tends to
infinity. This behavior is typical for all purely radial systems. Thus, the
conventional methods of stability theory cannot be applied to study ROI in
models with purely radial orbits. Suitable systems must have DFs with at least
small but finite angular momentum dispersion.

Note that the generalized polytropes are also unsuitable for studying the
instability by analytical methods (by solving the eigenvalue problem) because
of the singular behavior of the density and the potential at $s \approx 2$.

In a separate work we shall present results of our study of ROI for
families of models discussed above.
The present work can be considered as the first step in this direction.

\bigskip
\section*{Acknowledgments}
The authors thank the referee for providing several valuable suggestions for
presentation of the material, and Dr. Jimmy Philip for editing the original
version of the article that helped to improve its quality. This work was
supported by Sonderforschungsbereich SFB 881 'The Milky Way System' (subproject
A6) of the German Research Foundation (DFG) and by  Basic Research Program
OFN-17 ``The active processes in galactic and extragalactic objects'' of
Department of Physical Sciences of  RAS.

\subsection*{APPENDIX. Approximate analytical solution for
model $q=\frac{1}{2}$ with almost radial orbits, $L_T\ll 1$}

We saw in Sec. 3 that for $q\ge \frac{1}{2}$ there are no models with purely radial
orbits. Now we construct a physically appropriate solution on the boundary $q=\frac{1}{2}$ for
arbitrary small but finite $L_T$. From (\ref{density}) and (\ref{pe}) one obtains:
\[
 \Psi''+\frac{2}{r}\,\Psi'=- 3N
 \left\{\ba{ll}
  \dfrac{\Psi^2}{L_T^2}&\textrm{for } 0<r<r_1\ ,\\
&\\
\dfrac{\Psi}{r^2}-\dfrac{L_T^2}{4r^4}&\textrm{for } r_1<r<1\ . \ea \right.
 \eqno{({\rm A1})}
\]
In the above equation, $r_1$ defined by $L_T^2=2r_1^2\Psi(r_1)$ separates two
regions, I and II.
In general, there is a region III adjacent to the sphere boundary (see Sec. 2),
but for small
$L_T$ it can be ignored since its width is of the order of $L_T^2$.
Equation (A1) is to be solved with boundary conditions
\[
  \Psi(1;L_T,N)=0\,,\ \ \Psi'(1;L_T,N)=-1\,,\ \ \Psi'(0;L_T,N)=0\,.
\]

In region II (A1) is a inhomogeneous linear equation. The solution satisfying
the
boundary conditions at the right boundary $r=1$ is
\[
\Psi_{II}(r)=\frac{1+4\nu^2}{9+4\nu^2}\,\frac{L_T^2}{4}\Bigl[
  \frac{1}{r^2}-\frac{3}{2\,\nu\sqrt{r}}\,\sin\Bigl(\nu\,\ln\frac{1}{r}\Bigr)-\]\[-
  \frac{1}{\nu\sqrt{r}}\,\cos\Bigl(\nu\,\ln\frac{1}{r}\Bigr)\Bigr]+
  \frac{1}{\sqrt{r}}\,\sin
  \Bigl(\nu\,\ln\frac{1}{r}\Bigr)\ ,
   \eqno{({\rm A2})}
\]
where $\nu=\sqrt{3N-{\case{1}{4}}}$ is a {\it real} parameter.
Taking into account that for $L_T\ll 1$ radius
$r_1$ is also very small, $r_1\ll 1$, and ignoring trigonometric terms in
square brackets  in (A2), we obtain
 \[
L_T^2=4\,r_1^{3/2}\,\frac{\sin(\nu\,\Lambda_1)}{\nu}\,\frac{9+4\nu^2}{17+4\nu^2}\ ,\
\ \ \Lambda_1\equiv \ln\frac{1}{r_1}\ ,
 \eqno{({\rm A3})}
\]
\[
\Psi_{II}(r_1)=\frac{2\,(9+4\nu^2)}{17+4\nu^2}\,\frac{\sin\,(\nu\Lambda_1)}{\nu\,\sqrt{r_1}}\ .
 \eqno{({\rm A4})}
\]
Since the function $\Psi(r)$ is positive, the condition $\nu\Lambda_1<\pi$ must
be
satisfied.

\begin{figure}
\centerline {{\includegraphics [width=87mm] {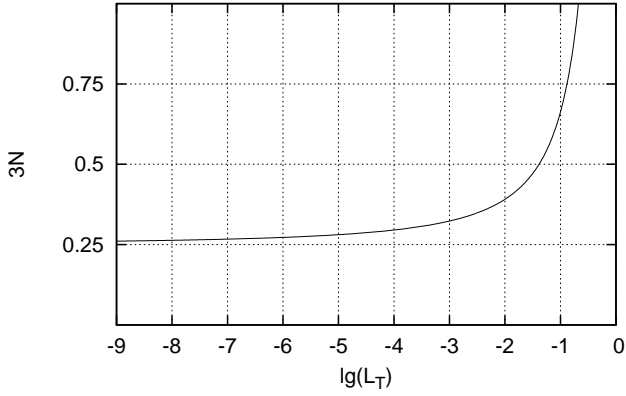}}}
\caption{\small The tripled normalization constant $N(L_T)$ for models
with $q=1/2$ and small $L_T$. }
 \label{fig:app}
\end{figure}

In the region I (A1) can be written using new independent variable $x\equiv {r}/{r_1}$:
\[
  \frac{1}{x^2}\frac{d}{dx}\,x^2\,\frac{d\Theta}{dx}=-{\case{3}{2}}\,N\,\Theta^2
  \equiv -{\case{1}{8}}\,(1 +4\nu^2)\,\Theta^2\ ,
  \eqno{({\rm A5})}
\]
where $\Theta(x)\equiv \Psi_I(r_1\,x)/\Psi_I(1)$ is a new unknown function. The
boundary conditions to be satisfied are:
 \[
\Theta'(0)=0\ ,\ \ \Theta(1)=1\ ,\]\vspace{-4mm}\[
\Theta'(1)=-\frac{21+20\nu^2}{4\,(9+4\nu^2)}-\frac{\nu\,(17+4\nu^2)}
{2\,(9+4\nu^2)}\, \cot\,(\nu\Lambda_1)\ .
 \eqno{({\rm A6})}
\]
The expression for $\Theta'(1)$ follows from the continuity of the first
derivative of the potential at $r=r_1$.  Equation (A6) with boundary conditions
(A7) can be solved numerically using standard shooting method for
$\nu_1\equiv\nu(\Lambda_1)$, where  $\Lambda_1$ is considered as a control
parameter. Then the relation $\nu=\nu(L_T)$ (and also
$3N(L_T)=\nu^2(L_T)+\frac{1}{4}$)  is obtained from the equality which follows
straightforwardly from (A3):
\[
  L_T=2
  \exp\,(-\case{3}{4}\,\Lambda_1)\,\sqrt{
  \dfrac{\sin\,(\nu_1\,\Lambda_1)}{\nu_1}\,\dfrac{9+4\nu_1^2} {17+4\nu_1^2}}\ ,
  \eqno{({\rm A7})}
\]
 The dependence of the tripled normalization
constant $N$ for small $L_T$ is shown in Fig. \ref{fig:app}. (Recall
that for $q=\frac{1}{2}$ we have  $ (L_T) _ {\textrm {iso}}
(\frac{1}{2}) = 0.6682$ and $3N\bigl((L_T) _ {\textrm {iso}}\bigr)
=4.686$.) It was useful to start the shooting procedure from
$\Lambda_1 \simeq 30$ which implies very small $r_1$ and
$3N-\frac{1}{4}\ll 1$. For large $\Lambda_1$, one can find an
asymptotic expansion for $\nu$  by applying perturbation theory to
(A6) and using factor $\frac{1}{8}$ in the r.h.s. as a small
parameter:
\[
  \nu \approx \frac{\pi}{\Lambda_1}-\kappa\,\frac{\pi}{\Lambda_1^2}\ ,
 \eqno{({\rm A8})}
\]
where $\kappa = 68/39$.  Note that from (A8) and (A3) it follows that
\[\frac{\sin(\nu\Lambda_1)}{\nu}\approx \kappa, \ \ L_T^2\approx
\frac{36}{17}\,\kappa\, r_1^{3/2}={\cal O}(r_1^{3/2}),\] i.e., $r_1={\cal
O}(L_T^{4/3})$, that justifies omitting trigonometric contributions of order
$L_T^2$ in derivation of (A3). For potential in the center we have an estimate
\[
 \Psi(0)\approx {\frac{49}{96}}\,
 \Big({\frac{36}{17}} \,\kappa\Big)^{4/3}\,L_T^{-2/3} = 2.91\,L_T^{-2/3}\ .
  \]
This analytical solution shows that the potential become singular with $L_T \to 0$,
but is remains regular as long as $L_T$ is arbitrary small, but finite.

\begin{thebibliography}{99}

\bibitem[\protect\citeauthoryear{Agekyan}{1962}]{Agk}
Agekyan T. A., 1962, Vestnik Leningrad. Gos. Univ., Ser. math., mech., astr.,
No 1, 152 (in Russian)

\bibitem[\protect\citeauthoryear{Antonov}{1973}]{Ant}
Antonov V. A., 1973, English translation in: de Zeeuw, T., ed. Proc. IAU Symp. 127,
Structure and Dynamics of Elliptical Galaxies, Reidel, Dordrecht, p. 549

\bibitem[\protect\citeauthoryear{\it Barnes, etc.}{1986}]{BGH}
Barnes J., Goodman J.,  Hut P., 1986, ApJ, 300, 112

\bibitem[\protect\citeauthoryear{\it Binney \& Tramain}{2008}]{BT}
Binney J.,  Tremain S., 2008, Galactic Dynamics: Second Edition. Princeton University Press,
Princeton, NJ, USA

\bibitem[\protect\citeauthoryear{\it Bisnovatyi-Kogan}{1969}]{BKZ}
Bisnovatyi-Kogan G. S., Zel'dovich Ya. B., 1969, Astrofizika, {5},
425 (in Russian)

\bibitem[\protect\citeauthoryear{BJ}{1968}]{BJ}
Bouvier P., Janin G., 1968, Publ. Obs. Gen$\grave{\rm e}$ve, {A74}, 186

\bibitem[\protect\citeauthoryear {\it Chandrasekhar}{1939}]{Chandr39}
Chandrasekhar C., 1939, An introduction to the study of stellar
structure. Dover publications, Inc.

\bibitem[\protect\citeauthoryear {\it Dejonghe}{1986}]{D86}
Dejonghe H., 1986, Phys. Rep., 133, 217

\bibitem[\protect\citeauthoryear {\it Dejonghe \& Merritt }{1992}]{DM92}
Dejonghe H., Merritt D., 1992, ApJ, 391, 531


\bibitem[\protect\citeauthoryear{\it Ernst \& Just}{2013}]{EJ13}
Ernst A., Just A., 2013, MNRAS, {429}, 2953

\bibitem[\protect\citeauthoryear{\it Fridman \& Polyachenko}{1984}]{FP84}
Fridman A. M., Polyachenko V. L., 1984, Physics of Gravitating Systems.
Springer, New York

\bibitem[\protect\citeauthoryear{\it Gelfand \& Shilov}{1964}]{GeSh}
Gelfand I. M., Shilov G. E., 1964, Generalized functions. Academic Press,
Inc.

\bibitem[\protect\citeauthoryear {\it Henon}{1973}]{Henon73}
 H\'{e}non M., 1973,  A\&A, {24}, 229

\bibitem[\protect\citeauthoryear{\it Kharchenko et al.}{2009}]{Kh09}
Kharchenko, N. V., Berczik, P., Petrov, M. I., Piskunov, A. E., R\"oser, S., Schilbach,
E., 2009,
A\&A 495, 807

\bibitem[\protect\citeauthoryear{\it Merritt}{1985}]{Merritt}
Merritt D., 1985, AJ, {90}, 1027

\bibitem[\protect\citeauthoryear{\it Osipkov}{1979}]{Osipkov}
 Osipkov L. P., 1979, Soviet Astron. Lett. {5}, 42

\bibitem[\protect\citeauthoryear{\it Palmer}{1994}]{Palmer}
Palmer P. L., 1994, Stability of collisionless stellar systems: mechanisms
for the dynamical structure of galaxies. Astrophysics and Space Science
Library, Kluwer, Dordrecht, Boston

\bibitem[\protect\citeauthoryear{\it Polyachenko et al.}{2010}]{PPS1}
Polyachenko V. L., Polyachenko  E. V., Shukhman I. G., 2010, Astron. Lett.,
{36}, 175

\bibitem[\protect\citeauthoryear{\it Polyachenko et al.}{2011}]{PPS2}
Polyachenko E. V., Polyachenko  V. L., Shukhman I. G., 2011, MNRAS, {416}, 1836

\bibitem[\protect\citeauthoryear {\it Richstone \& Tremaine}{1984}]{RT84}
Richstone D.,  Tremaine S., 1984, ApJ, {286}, 27

\bibitem[\protect\citeauthoryear {\it Touma, Tremaine}{1997}]{RT97}
Touma J.,  Tremaine S., 1997, MNRAS, {292}, 909


\end{thebibliography}
\end{document}